\documentclass[]{emulateapj}
\usepackage{graphicx,amsmath,rotating}

\begin{document}

\title{Pencil-Beam Surveys for Trans-Neptunian Objects:\\Novel Methods for Optimization and Characterization}
\author{\bf Alex H. Parker}
\affil{\textit{Department of Astronomy, University of Victoria}}
\email{alexhp@uvic.ca}
\and
\author{\bf JJ Kavelaars}
\affil{\textit{Herzberg Institute of Astrophysics, National Research Council of Canada}\\}

\slugcomment{Accepted for publication in PASP March 1, 2010.}

\begin{abstract}
Digital co-addition of astronomical images is a common technique for increasing signal-to-noise and image depth. A modification of this simple technique has been applied to the detection of minor bodies in the Solar System: first stationary objects are removed through the subtraction of a high-SN template image, then the sky motion of the Solar System bodies of interest is predicted and compensated for by shifting pixels in software prior to the co-addition step. This ``shift-and-stack'' approach has been applied with great success in directed surveys for minor Solar System bodies. In these surveys, the shifts have been parameterized in a variety of ways. However, these parameterizations have not been optimized and in most cases cannot be effectively applied to data sets with long observation arcs due to objects' real trajectories diverging from linear tracks on the sky. This paper presents two novel probabilistic approaches for determining a near-optimum set of shift-vectors to apply to any image set given a desired region of orbital space to search. The first method is designed for short observational arcs, and the second for observational arcs long enough to require non-linear shift-vectors. Using these techniques and other optimizations, we derive optimized grids for previous surveys that have used ``shift-and-stack'' approaches to illustrate the improvements that can be made with our method, and at the same time derive new limits on the range of orbital parameters these surveys searched. We conclude with a simulation of a future applications for this approach with LSST, and show that combining multiple nights of data from such next-generation facilities is within the realm of computational feasibility.
\end{abstract}

\shorttitle{Optimizing Pencil-Beam Surveys for TNOs}
\shortauthors{Parker \& Kavelaars}
\maketitle

\section{Introduction}

Detecting faint objects in the Solar System is a technically challenging problem. The approaches taken in the past can be broken into two basic categories: inter-frame source detection and linking, and the ``shift-and-stack'' approach. Inter-frame detection, where transient sources are detected in individual exposures and motion is linked between multiple exposures, is widely employed and favored in planned proposed surveys like Pan-STARRS (Denneau et al. 2007) and LSST (Axelrod et al. 2009) due to its simplicity and robustness. This method, however, does not exploit imaging data to the fullest extent possible, since detections are subject to the completeness limit of individual frames. 

Since Solar System objects are moving across the sky, their images trail and the noise contribution from the sky background increases. Trailing losses create an upper limit to individual frame exposure times for efficient detection of Solar System objects. It is advantageous to combine multiple exposures, each short enough that trailing effects are negligible. Co-addition is routinely done to improve the SN of stationary sources and to remove cosmic ray events, but in order to apply it to the detection of moving objects several other steps must be performed. First, contamination from stationary sources can be removed by subtracting a high-SN template image from each image in the set. Next, the sky motion of the objects of interest must be predicted and this motion compensated for by shifting each image's pixels in software. When the images are then combined only the flux from sources that moved at the predicted rates will be constructively added.

This method first appeared in literature in Tyson et al. (1992), though details of the application were limited. An early detailed description of this technique was presented by Cochran et al. (1995). A set of images from the \textit{Hubble Space Telescope} (HST) \textit{Wide Field Planetary Camera 2} (WFPC2) were combined to allow statistical detection of very faint TNOs. The sky motion of the sources was parameterized as angular rates ($\dot{\theta}$) and angles on the sky ($\phi$). Gladman et al. (1997) used the same technique in order to combine a series of images from the \textit{Palomar} 5m telescope in the hope of constraining the size distribution of Kuiper Belt Objects (KBOs) by discovering objects with smaller radii than had been detected in most previous wide-area surveys that employed inter-frame detection. This survey was followed by further searches from 
\textit{Keck} (Luu \& Jewitt 1998; Chaing \& Brown 1999), 
\textit{CTIO} (Allen et al. 2001 and 2002; Fraser et al. 2008),
\textit{Palomar} (Gladman et al. 1998),
\textit{VLT} and \textit{CFHT} (Gladman et al. 2001), 
which employed the same technique to varying degrees of success. 
Recently, Fraser \& Kavelaars (2009) and Fuentes, George, \& Holman (2009) both used similar techniques to search \textit{Subaru} data for KBOs as faint as $R \sim 27$.


All of these surveys had observational arcs less than two days in length. Over periods this short, the motion of outer Solar System sources can be well approximated (with respect to ground-based seeing) by the linear trajectories that the method these surveys implemented implicitly assumes. For longer arcs or higher-resolution data, however, the approximation of a linear trajectory is no longer adequate. Bernstein et al. (2004) employed a more advanced approach, which they called ``digital tracking,'' to extremely high-resolution HST \textit{Advanced Camera for Surveys} (ACS) data. Even in arcs on the order of a day in length, nonlinear components of motion were non-negligible relative to the ACS PSF. In order to stack frames such that distant moving sources add constructively in this regime it becomes necessary to generate nonlinear shift-vectors via full ephemerides from orbits of interest. They stacked arcs up to 24 hours in length using shift-vectors parameterized as a grid over three parameters: two linear components of motion and heliocentric distance $d$. By sampling densely over this grid, they were able to recover orbits with $25$ AU $\leq d \leq \infty$ and $i < 45^{\circ}$, and detect objects as faint as $R\sim28.5$. However, this dense sampling, somewhat unconstrained discovery volume, and extremely small PSF required $\sim7\times10^{5}$ shift-vectors, which translated into $\sim10^{14}$ pixels to be searched. Bernstein et al. found that stacking more than 24 hours of their data became computationally prohibitive. 

In the age of ground-based gigapixel CCD mosaic imagers, how is data obtained over long arcs going to be efficiently exploited to search for faint sources? In the interest of exploring massive datasets with long observation arcs to as deep a limit as possible, we have created a simple probabilistic method for generating an optimized set of shift-vectors for any imaging survey and any interesting volume of orbital element space. By generating shift-vectors in a probabilistic way, exploring an explicitly demarcated region of orbital parameter space becomes straightforward. The targeted region of orbital parameter space is searched with uniform sensitivity, simplifying the accurate characterization of any detected sample of objects. This method also avoids searching non-physical or uninteresting parameter space, thereby maximizing scientific return for minimum computational cost.

In Section \ref{MTE}, we define the maximum tracking error, the basic parameter that defines the density of a grid. Section \ref{MC} describes previous analytical methods for search-grid parameterization, our probabilistic approach, and a grid-tree structure for optimizing multiple-night arcs. Section \ref{COMP} compares the efficiency of our method to those used in previous surveys, and in Section \ref{LSST} we show an example application of this method to the kind of data that may be acquired by the Large Synoptic Survey Telescope (LSST).

\section{Generating Search Grids} \label{MC}

\subsection{Estimates of On-Sky Motion}\label{AM}

Previous on-ecliptic surveys (eg., Fraser \& Kavelaars 2009) have used simple analytical estimates for the apparent on-sky rates of motion for distant Solar System objects in order to generate shift rates. For an on-ecliptic observation of a field $\beta$ degrees from opposition, the range of on-sky angular rates of motion $\dot{\theta}$ and angle from the ecliptic $\phi$ for a distant object at heliocentric distance $d$ and geocentric distance $\Delta$ with inclination $i$ and eccentricity $e$ can be approximated by the vector addition of the reflex motion of the object due to Earth's motion (inversely proportional to $\Delta$) and the intrinsic on-sky orbital motion of the object by Kepler's law (proportional to $d^{-3/2}$):


\begin{equation}
\begin{split}
\dot{\theta} \simeq & 148\! \biggl[  \left( \cos(\beta)(\Delta)^{-1} - v_{d}d^{-\frac{3}{2} }  \cos(i) \right)^2\\
& + \left( v_{d}d^{-\frac{3}{2}}  \sin(i) \right)^2 \biggr]^{\frac{1}{2}} \quad \arcsec \: hr^{-1} \label{rate}
\end{split}
\end{equation}

\begin{equation}
\phi \simeq  \arcsin\left( 148 v_{d} \dot{\theta}^{-1} d^{-\frac{3}{2}}\sin(i) \right) \label{angle} 
\end{equation}

where $v_{d} = \sqrt{1 \pm e}$, representing the fraction of the mean angular velocity of an orbit at pericenter ($+e$) or apocenter ($-e$), compared to a circular orbit at heliocentric distance $d$. 

This motion can also be parameterized as two components of angular rates of motion, one parallel and one perpendicular to the ecliptic. Re-arranging Eqn. \ref{rate} into parallel and perpendicular components, we find:
\begin{eqnarray}
\dot{\theta}_\shortparallel \simeq & 148\! \left( \cos(\beta)(\Delta)^{-1} - v_{d}d^{-\frac{3}{2} }  \cos(i) \right) \, \arcsec hr^{-1} \label{thetapar}\\
\dot{\theta}_\perp \simeq & 148 v_{d}d^{-\frac{3}{2} }  \sin(i)  \: \arcsec hr^{-1} \label{thetaperp}
\end{eqnarray}

The $\dot{\theta}$ and $\phi$ parameterization is convenient for visualization purposes, but in Section \ref{MTE} we show that the $\dot{\theta}_\shortparallel$ and $\dot{\theta}_\perp$ parameterization is more appropriate for generating a search grid.

The angular rate of motion is lowest (for a given $d$ and $\Delta \simeq d$) for $i=0^\circ$ orbits at pericenter, since $v_{d}$ is at its highest and the parallax and orbital motion vectors are anti-aligned.\footnote{This assumes that the object's distance and eccentricity are not such that that $v_{d}/d^{\frac{3}{2}} \geq v_{\oplus}$ - in other words, the object is not overtaking Earth at opposition. If this were the case, $\dot{\theta}$ is lowest at $i = 180^\circ$.} The highest angular rate of motion at opposition is reached in two cases, depending on the maximum inclination and eccentricity considered ($i_{max}$ and $e_{max}$):

\begin{eqnarray*}
\dot{\theta} \mbox{ max when}
\begin{cases} 
i = 0^\circ, \, \textrm{apocenter}, &\mbox{if  } i_{max} < i_0 \\
i = i_{max}, \, \textrm{pericenter},  &\mbox{if  } i_{max} > i_0,
\end{cases}
\end{eqnarray*}

where 

\begin{equation*}
i_0 = \arccos\!\left( \frac{v_q^2 - v_Q^2}{2\sqrt{d}\cos(\beta)v_q } + \frac{v_Q}{v_q} \right)
\end{equation*}

and $v_{q} = \sqrt{1+e_{max}}$, $v_{Q} = \sqrt{1-e_{max}}$.

Given this approximation, the highest $\phi$ any orbit can achieve is a strong function of $d$. If we consider an orbit with $e\rightarrow1$ at pericenter with $i = 90^\circ$, and approximate $\Delta \sim d$, then by Eqn. \ref{angle} it can be shown that:

\begin{eqnarray}
\phi_{max} \simeq  &\pm  \arcsin\left( \sqrt{\frac{2}{\cos^2(\beta)d + 2}} \right) \label{maxangle}
\end{eqnarray}

\subsection{Grid Geometry and Maximum Tracking Error $\epsilon_{max}$} \label{MTE}

In using digital tracking, the motions of real sources are approximated by a grid or some distribution of shift- or motion-vectors. In order to quantize the effectiveness of this strategy, we define a maximum tracking error $\epsilon$ to be the largest possible error tolerated between any real object's motion and the nearest predicted motion vector. To estimate its effect, we compare the final signal-to-noise ratio $S/N'$ for a faint circular source with flux $f$ and area $A_{0}$ before and after a linear tracking error $\epsilon$ is added in an image with exposure time $t$. The linear tracking error $\epsilon$ adds an additional rectangular region to the area of the source, with width $2R$ (where $R$ is the aperture radius applied to the initial circular source) and length $\epsilon$. This rectangle has area $2R\epsilon$, and the total area of the blurred source is now: 

\begin{equation*}
A' = \pi R^{2} + 2R\epsilon = A_{0}\left(1 + \frac{2}{\pi R}\epsilon\right). 
\end{equation*}

The final signal-to-noise $S/N'$ is given by:
\begin{equation}
\begin{split}
\frac{S}{N'} &\simeq \frac{ f \sqrt{t} }{\sqrt{f_{sky} A_{0}(1 + \frac{2}{\pi R} \epsilon)}} \\
& =\frac{S}{N_{0}}\frac{1}{\sqrt{1 + \frac{2}{\pi R} \epsilon}} \quad \mbox{for $f \ll f_{sky}$}
\end{split}
 \end{equation}
So, for a search-grid with maximum tracking error $\epsilon_{max}$, the worst possible signal-to-noise degradation is a factor of $F = (1 + \frac{2}{\pi R} \epsilon)^{-\frac{1}{2}}$. Given a gaussian point-spread function with a full-width at half-maximum of $\Gamma$, the radius at which the signal-to-noise of a sky-dominated source is maximized is $R \simeq 0.68 \Gamma$. If we define this as our initial aperture radius, then the maximum allowable tracking error given a desired limit on $F$ becomes 

\begin{equation} \label{SNdeg}
\epsilon_{max} \: \simeq \: \frac{\pi}{2} 0.68 \Gamma (F^{-2} - 1). 
\end{equation}

Previous surveys have used widely varying values for $\epsilon_{max}$ with respect to the typical seeing of that survey, ranging from $\sim0.1\Gamma - 3\Gamma$. Surveys that searched data by eye were limited in the number of shift-vectors they could apply, and leveraged the eye's robust noise discrimination to compensate for the loss in signal-to-noise due to using large $\epsilon_{max}$ (eg., Fraser \& Kavelaars 2009). Surveys using automated detection pipelines have decreased $\epsilon_{max}$ to limit their signal-to-noise loss, and leveraged massive computational resources to compensate for the increased number of required shift-vectors (eg., Fuentes et al. 2009).

In order to evaluate the maximum tracking error of any search-grid of shift-vectors applied to a given data set, we determine the distribution of shifts from the initial image to the final image, where the shift for orbit $k$ in image $i$ is:
\begin{equation}\label{svec}
\begin{split}
 {\bf v}_{k,i} &= [ d\alpha_{k,i},\, d\delta_{k,i} ]  \\
 &= \left[\cos(\delta_{k,0})(\alpha_{k,i} - \alpha_{k,0}),\, (\delta_{k,i} - \delta_{k,0})\right]  .
\end{split}
 \end{equation}
 
The maximum separation any real orbit's motion from one of these offsets determines the maximum tracking error of the search-grid. After time $t$ has elapsed, the on-sky angular separation $\Omega$ of two objects starting from the same initial position with identical angular rates of motion $\dot{\theta}$ and on-sky angles of motion separated by small angle $d\phi$ is given by 

\begin{equation}
\Omega = 2 \dot{\theta} t \sin\left(\frac{d\phi}{2}\right) . \label{delta}
\end{equation}

The maximum tracking error between a search-grid of shift-vectors and a real distribution of orbits depends on the geometry of the search-grid. Consider the grid spacing of a geometrically regular search-grid in the plane defined by the total changes in RA and DEC made by moving sources during the observations, which we will refer to as the ($d\alpha$, $d\delta$) plane. If the points of the search-grid are arrayed at the vertices of identical adjoining geometric cells on the ($d\alpha$, $d\delta$) plane, the point most distant from any point on the search-grid is the point at the center of each cell. The distance from any vertex of a cell to the center of that cell defines the maximum tracking error of that search-grid.

\begin{figure*}
\begin{centering}
\includegraphics[width=5cm]{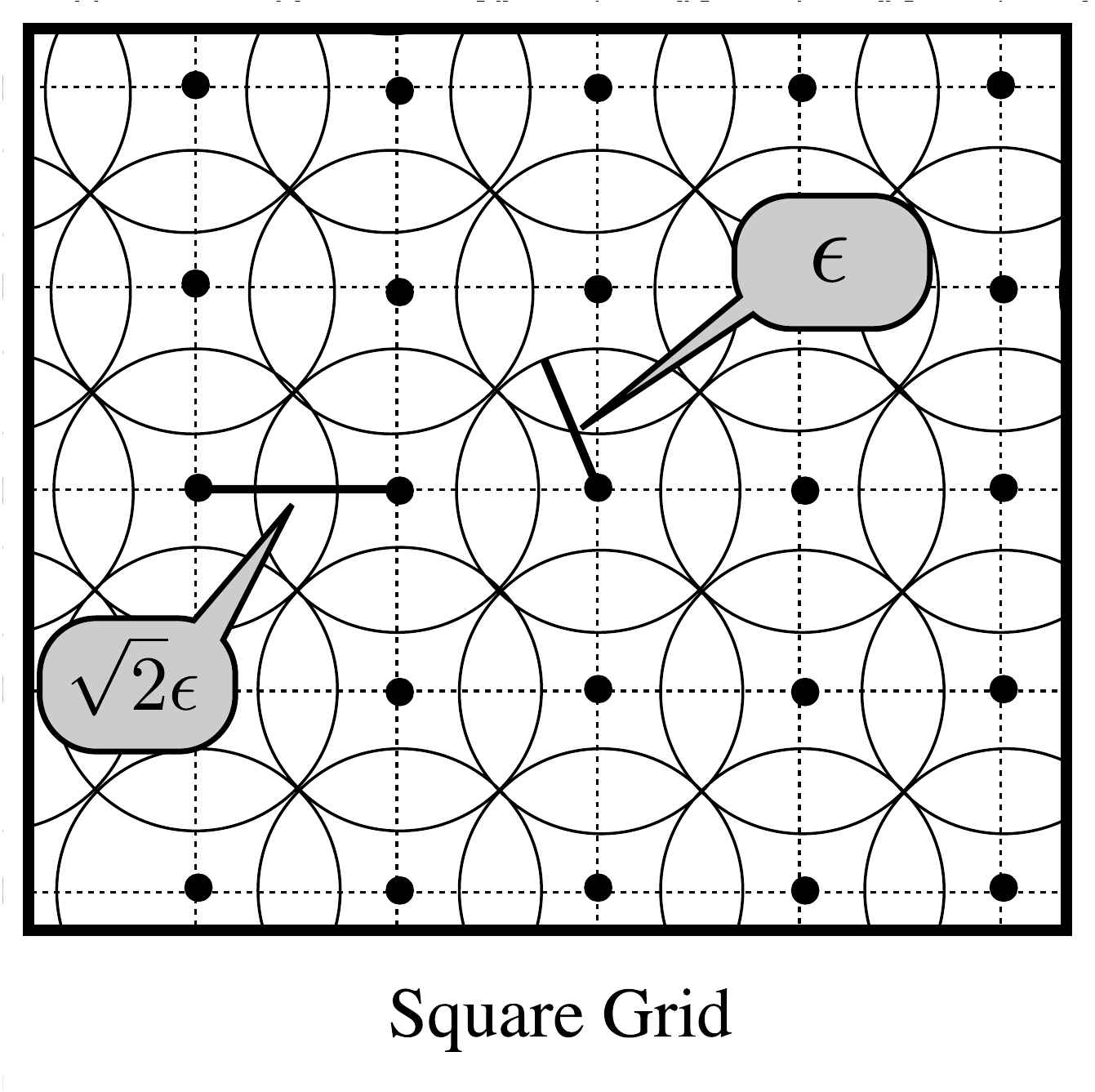}
\includegraphics[width=5cm]{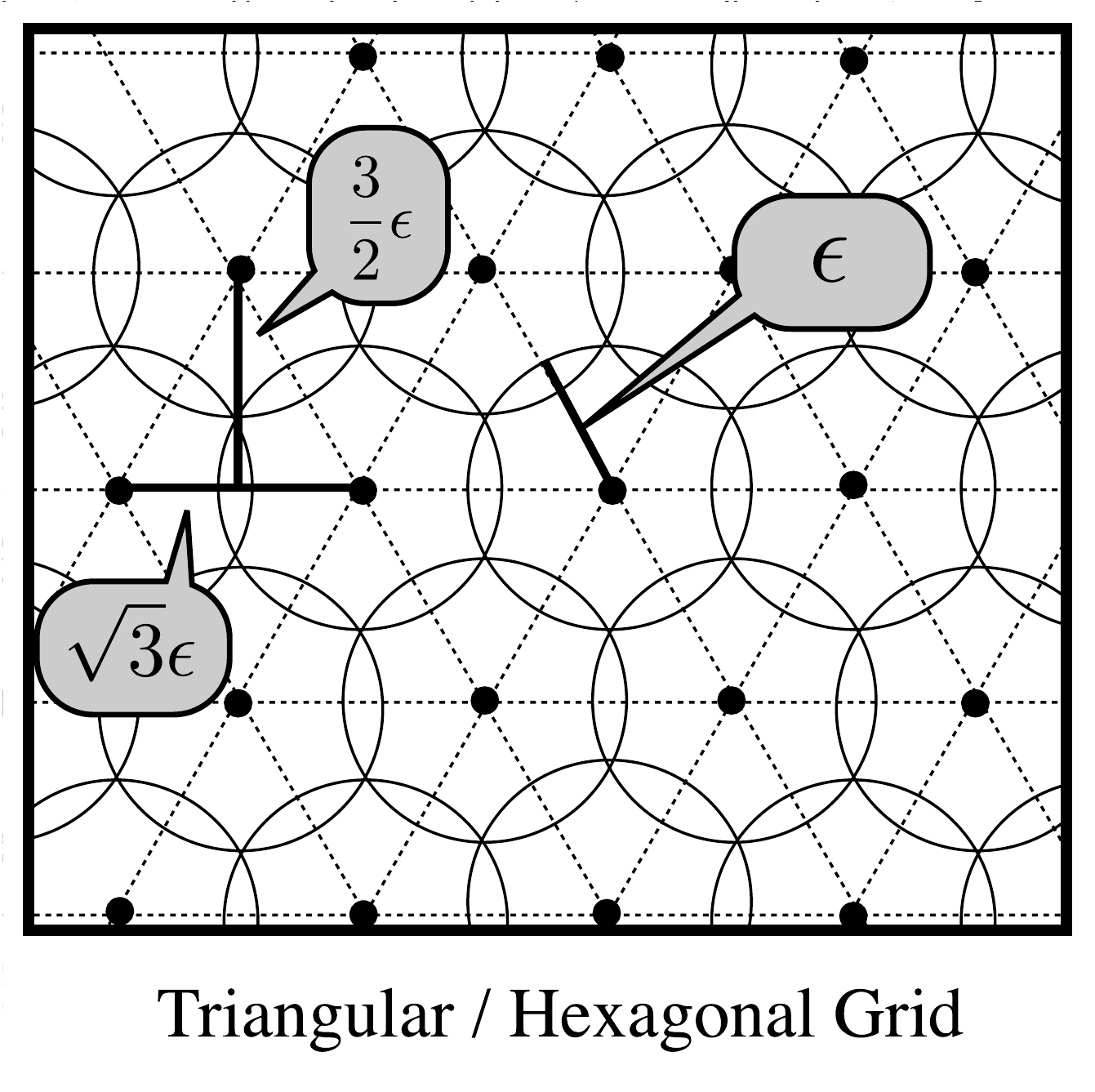}
\caption{Comparison of grid geometries. Left panel: Packing of a square lattice, with scales illustrated. Right panel: packing of a hexagonal lattice, with scales illustrated.}
\label{grid_geom}
\end{centering}
\end{figure*}

For points arrayed on the vertices of a grid of adjoining rectangles with sidelengths $dA$ and $dB$, the maximum tracking error is $\epsilon_{max} = \frac{1}{2}\sqrt{dA^2 + dB^2}$. In the case there $dA = dB$, this becomes $\epsilon_{max} = \sqrt{2}dA/2$. This geometry well approximates the search-grids used by most previous surveys. The left panel of Figure \ref{grid_geom} illustrates a grid with this geometry.

If a fixed ($d\dot{\theta}$,$d\phi$) grid spacing is adopted such that a maximum tracking error $\epsilon_{max}$ is satisfied after time $t$ for objects with angular rate of motion $\dot{\theta} = \dot{\theta}_{0}$, objects with $\dot{\theta} > \dot{\theta}_{0}$ will see increasing tracking errors. To illustrate, consider a roughly rectangular grid where $dA\simeq d\dot{\theta}t$ and $dB\simeq 2 \dot{\theta} t \sin(d\phi/2)$. Inserting these values into our definition of $\epsilon_{max}$ for a rectangular grid, we see that $\epsilon_{max}\propto\dot{\theta}$. A fixed ($d\dot{\theta}$,$d\phi$) grid parameterization over-samples motions with low $\dot{\theta}$, and under-samples motions with high $\dot{\theta}$. 

The over- and under-sampling problem can be remedied by parameterizing the sky motion as two components of angular rates of motion, one parallel and one perpendicular to the ecliptic ($\dot{\theta}_\shortparallel$ and $\dot{\theta}_\perp$). By adopting a fixed spacing in ($\dot{\theta}_\shortparallel,\,\dot{\theta}_\perp$) along with angles from the ecliptic ($\phi_{min},\,\phi_{max}$) between which motions are confined, a uniform $\epsilon_{max}$ can be maintained for all angular rates of motion. This is the approach that was adopted by Fuentes et al. (2009).

The packing efficiency of a search-grid can be improved over the regular rectangular case. If the area of ($d\alpha$, $d\delta$) to be searched by a grid is much larger than the unit of area searched by an individual grid point such that we can largely ignore boundary conditions, the proof by Gauss (1831) regarding the most efficient regular packing of circles on a plane holds. In this case, defining the points of the search-grid as the vertices of a grid of adjoining equilateral triangles will produce the highest packing efficiency. Considering a grid of equilateral triangles with sidelength $dA$, the maximum tracking error is $\epsilon_{max} = dA/\sqrt{3}$ (where the effective $dB$ spacing becomes $3\epsilon_{max}/2$). The right panel of Figure \ref{grid_geom} illustrates a grid with this geometry. Both our short- and long-arc solutions described in the next section produce a grid with approximately this geometry, and typically require $\geq 20\%$ fewer grid points than a similarly-defined rectangular grid with the same $\epsilon_{max}$.

Regardless of the grid geometry, each cell is basically covering a small area of the final ($d\alpha$, $d\delta$) plane. The final area of the ($d\alpha$, $d\delta$) plane covered by real objects is approximately $\propto t^2$ (Equations \ref{thetapar} and \ref{thetaperp}, neglecting acceleration), while the area covered by a single grid point is $\propto \epsilon_{max}^2$. The total number of grid points $N$ required to search a given population scales as $t^2 / \epsilon_{max}^2$.

\subsection{Probabilistic Methods for Generating Shift-Vectors}

The methods for generating shift-vectors described above are cumbersome to analytically generalize over any region of the sky, arbitrary regions of orbital parameter space, and arbitrarily long observational baselines as they still suffer from the problem of non-linear components of motion that become significant as $t$ grows. In this regime it becomes necessary to add additional dimensions to the search grid ($\ddot{\theta}$ and $\dot{\phi}$, for example), and selecting an optimized set of grid-points becomes a challenge. Instead of pursuing an analytical solution, in the following section we outline a simple probabilistic approach for defining an optimum set of shift-vectors for any region of the sky and over any length of observational baseline at any solar elongation.

Instead of approximating the minimum and maximum angular rates of motion and angle on the sky for an ad-hoc subset of the orbits of interest, our method takes the following approach:  First, generate a very large sample of synthetic orbits that fill (in a characterized way) the orbital parameter space of interest. Second, use ephemeris software to determine the ($\alpha,\, \delta$) shift between frames for every synthetic orbit in every imaged epoch. Finally, search for the optimum (smallest) subset of orbits whose motions accurately represent the motions of  the entire set of synthetic orbits within a maximum tracking error tolerance. The set of ($\alpha,\, \delta$) shifts (translated into ($x,\,y$) pixel shifts) from this representative set of orbits is our optimum set of shift-vectors. A more detailed description of the method follows.

\subsubsection{Initial sample generation}
To create our initial sample, we use a randomly generated set of synthetic orbits that fill a region of orbital parameter space of interest such that the orbital ephemerides intersect the imaged field during the times of observation. The approach to filling orbital parameter space is subject to some consideration, as any orbital-space biases may translate into a skewed on-sky motion distribution, resulting in a biased selection of ``optimum'' orbits. Over small ranges of $d$, a uniform sampling is adequate, but for larger ranges we sample uniformly with respect to $d^{-c}$, where the index $c = 2$ is appropriate to generate a uniform distribution in $\dot{\theta}$ for observations near opposition. Once we have selected a given $d$, we select $a$, and $e$ from a uniform distribution between the minimum and maximum values set for those parameters, in the following order: first select $e$, then $a$ such that it falls within its limits and also $a \geq d(1-e)$. 

In cases where a population has a particular orbital parameter that is well-constrained, it is preferable to generate this parameter first (eg., $a$ for a mean-motion resonance), and then force the other parameters onto as uniform a grid as possible. In the case of populations in mean-motion resonances, $a$ is pinned at the resonant value $a_{mmr}$, then $d$ is selected from a uniform distribution between $a_{mmr}(1-e_{max})$ and $a_{mmr}(1+e_{max})$. Finally, $e$ is selected such that $a_{mmr}(1-e) \leq d \leq a_{mmr}(1+e)$. In either case, $a$, $e$, and $d$ define two possible values for $M$, which are given equal probability.

Sampling inclination from the distribution of an isotropic sphere (probability of $i \propto \sin(i)$) may seem appropriate so as not to bias the on-sky motion distribution to that of low inclination orbits, but because the component of motion perpendicular to the ecliptic is roughly $\propto \sin(i)$, and because the small area of any imaged field with respect to the entire sky places stronger limits on the phase-space volume available to high-inclination objects than that available to low-inclination objects, we contend that is preferable to sample $i$ from a uniform distribution. The minimum inclination orbit possible to be observed is $i_{min} = \arccos(l_{min})(1 - \frac{1}{d})$, where $l_{min}$ is the minimum ecliptic latitude of the field. In the case where retrograde orbits are of interest, duplicate the set of inclinations with $i_{r} = 180 - i$.

The remaining orbital elements to be generated are the node and the argument of perihelion. These can be rotated to force the object to fall on the imaged field during the dates of observation, and should fill the parameter space set by these constraints smoothly. $M$ and the argument of perihelion are coupled for objects in mean motion resonances, but  for the purposes presented here it is convenient to simplify the treatment by treating the two as independent.

Depending on the size of the orbital region of interest and the length of the observational arc, the number of synthetic orbits required to produce a statistically adequate sample varies. The basic requirement is that the density of points in the final ($d\alpha$, $d\delta$) plane is such that the max spacing between any two nearest neighbors in the initial sample is $\ll \epsilon_{max}$, ensuring that there are no empty bins in ($d\alpha$, $d\delta$) given circular bins with radius $\epsilon_{max}$. In trials, we found that this required the initial sample size to range from several thousand to hundreds of thousands of synthetic orbits. 

Once a set of synthetic orbits has been created, the shift-vectors for each orbit can be determined. We define the shift-vector for orbit $k$ in image $i$ to be a point in the ($d\alpha_i,\,d\delta_i$) as given by ${\bf v}_{k,i} $ in Eqn. \ref{svec}. The list of shift-vectors $[{\bf v}_{k,0...n}]$ is stored as a list linked to the orbit $k$ for which they were generated. We then compare each list of shift-vectors to determine the optimum set, which is the smallest subset which have motions that accurately represent the whole set within some spatial tolerance in the image plane. We set this tolerance by specifying that all synthetic orbits in the initial sample must be matched to the motions of at least one orbit in the optimum sample to within our maximum tracking error tolerance $\epsilon_{max}$ in every imaged epoch. In other words, every synthetic orbit $k$ must be matched to at least one optimum orbit $h$ in every image $i$ by $| {\bf v}_{k,i} -  {\bf v}_{h,i} | \leq \epsilon_{max}$.

\subsubsection{Short-arc solution}
\label{shortarc}

For observational baselines over which any nonlinear component of motion is negligible compared to $e$, the largest offset between any orbit and the nearest predicted motion vector will always occur in the final image $n$. To generate shift-vectors in this regime, we use the following approach:
\begin{enumerate}
\item Propagate all synthetic orbits in the initial sample forward to the final image ($n$) plane and record the final shift-vector ${\bf v}_{k,n}$ for each orbit $k$. Find the median of the distribution of shift-vectors and set the corresponding point in ($d\alpha_n,\,d\delta_n$) to be the center of a geometrically-optimum triangular grid with sidelength $dA = \sqrt{3}\epsilon_{max}$ that extends over a region of ($d\alpha_n,\,d\delta_n$) significantly larger than the that occupied by any synthetic orbits in the initial sample. 

\item From this initial large grid, select only those grid points that have one or more synthetic orbits whose final shift-vector lies within $\epsilon$ of it in ($d\alpha_n,\,d\delta_n$). Record the list of orbits that are matched to each grid point.

\item After the initial grid has been cut down to just those grid points that are matched to at least one synthetic orbit, find the smallest subset that are matched to \textit{unique} synthetic orbits: First, select the grid point that is matched to the largest number of synthetic orbits, set it and its list of orbits aside, then remove the references to each synthetic orbit recovered by that grid point from all other grid points. Repeat this process for the remaining grid points, until all the original synthetic orbits are accounted for in the set-aside list. All set-aside grid points are now identified as the optimum set for recovering the initial sample of synthetic orbits, given this grid geometry and orientation.

\item Iterate Steps 1-3 several times, each time rotating the initial grid around its central point by some small angle $d\mu$ (up to a total rotation of $60^\circ$). Record the number of optimum grid points required for each orientation, and select the orientation that requires the fewest. As a small refinement, if any of the optimum grid points are found to lie with \textit{centers} outside the distribution of synthetic orbits in the ($d\alpha_n,\,d\delta_n$) plane, we shift its center to the nearest ($d\alpha_n,\,d\delta_n$) point from a synthetic orbit. After this refinement, the set of optimum grid points from this orientation are defined as our final optimum set.

\end{enumerate}

These grid points are points in the final ($d\alpha,\,d\delta$) plane. We can back out the implied set of optimum ($\dot{\theta}_\alpha, \, \dot{\theta}_\delta$) rates by simply dividing the respective components of each grid point by the total observational baseline $t_b$.

The rotation in step 4 is included to address boundary conditions. If the region of ($d\alpha$, $d\delta$) searched by a grid is not $ \gg \epsilon_{max}$, then there will be a preferred orientation of the grid. However, if $\epsilon_{max}$ is small relative to the total region, this step is unnecessary.

If, instead of trimming the search-grid to as small a number of points as possible, it is preferable to create a ``safe'' grid that over-searches the boundaries in rate-space by some comfortable margin, then modify step 2 to select grid points that lie within a larger distance $N\times\epsilon_{max}$ of any final shift-vector. Step 3 must be skipped in this case. This maintains the grid spacing and geometry, but adds a ``buffer'' region around each synthetic orbit.

\subsubsection{Long-arc solution}
\label{longarc}

In the case that the observational baseline $t_b$ is long enough such that non-linear components of motion are non-negligible with respect to $\epsilon$, a different treatment is required. Instead of presuming that the largest offset between the motions of a given orbit and the nearest grid point will occur in the final frame $n$, we must instead ensure that for every synthetic orbit there is one non-linear shift-vector which matches its motions within $\epsilon$ \textit{in every frame}. While nonlinear components of motion are non-negligible, the dominant factor driving the number of shift-vectors required is still the linear components of motion.

\begin{enumerate}

\item After creating an initial sample of synthetic orbits, generate a search grid via the short-arc (linear) solution for the observations. When generating the search grid, set the effective maximum tracking error to be slightly smaller than the desired final maximum tracking error. This is to take into account that the final grid will be selected from a \textit{discrete} distribution; namely, the motions of the synthetic orbits in the initial sample. 

\item For each grid point $p$ on the optimum \textit{linear} search grid, take the list $l_{p}$ of orbits uniquely matched to that grid point and consider their shift-vectors in every imaged epoch. For each orbit $k$ in list $l_{p}$, find all the other orbits within $l_{p}$ that have motions that are matched to the motion of $k$ within the desired maximum tracking error $\epsilon_{max}$ in every frame. 

\item Select the single synthetic orbit $k_{1}$ with motions matched to the largest number of unique synthetic orbits in $l_{p}$ and set it aside as an \textit{optimum} orbit. 

\item If some of the orbits linked to the grid point $p$ are not recovered by the first optimum orbit $k_{1}$, remove all orbits recovered by the orbit $k_{1}$ from consideration, then repeat Step 3 to select a second optimum orbit $k_{2}$. 

\item Repeat Step 4 until all the orbits linked to the grid point $p$ are accounted for by one or more optimum orbits.

\item Repeat Steps $2-5$ for each grid point in the initial solution.

\end{enumerate}

The shift-vectors of the final optimum set of orbits are then identified as the optimum set of non-linear shift-vectors to apply to the given data set.

\subsubsection{Grid Tree: Optimization for multi-night arcs}
\label{alias}

Other significant optimizations can be made to reduce the total number of pixel additions required to search a given data set. One particular optimization that our probabilistic method lends itself to is the use of a multi-level grid tree described by Allen (2002). This approach recognizes that for multi-night stacks, there are essentially two distinct timescales: the unit time periods over which images are obtained (length of a single night), and the total observational baseline. Allen (2002) points out that a significant reduction in total number of required shift-vectors can be obtained if, instead of combining all images obtained over several nights with one search-grid, images are combined on a per-night basis (with a single-night grid), then these stacks are combined with a \textit{second} search grid to remove the motion that would have occurred over the entire period. If all else remains equal, this tree-of-grids structure results in fewer overall pixel additions, as redundant combinations of images taken on a single night are only performed once. Since this method relies on combining images that are the combination of other images, it can only be applied for combination algorithms that can nest: for example, weighted averaging and co-adding are usable, but not medians.

It is important to note that in combining a tree of grids, the resulting maximum tracking error $\epsilon_{max}$ goes as the sum of each level's tracking errors $\sum \epsilon_i$. As such, each level's tracking error must be limited to $\epsilon_i = \epsilon_{max} / N$, where $N$ is the total number of levels. In the case described here, the tree has two levels, and so the effective tracking error in each level must be limited to $\epsilon_i = \epsilon_{max} / 2$. Thus, while this 2-level case requires fewer pixel additions, it results in significantly more pixels to \textit{search} than in the 1-level case.

While we have not implemented this method, we can estimate its results. Since the number of grid points $N$ is approximately proportional to $t^2 / \epsilon^2$, and the 2-level case requires $\epsilon_{i} = \epsilon_{max} / 2$, we can estimate that this method will require four times as many shift-vectors per unit time interval as a 1-level grid. If we consider $n$ nights of data, where the length of a night is equal to the length of a day such that the total observational baseline is $t_{b} = (2n - 1)$ in units of the length of a single night. Thus, if the 1-level grid requires $N_{1}$ shift-vectors for the full observational timeline, we can estimate that the 2-level grid for the full length of time will require $N_{2,full} \simeq 4N_{1}$. The 2-level method will require $\sim4$ times more pixels be searched in the final stacks than the 1-layer method. However, to determine the improvement in pixel additions, we need to determine the size of each nightly grid:

\begin{equation}
N_{2,nightly} \simeq \frac{N_{2,full}}{t_{b}^2} \simeq \frac{4N_{1}}{(2n-1)^2}
\end{equation}

For surveys where the number of observations taken per night $N_{obs} \gg N_{2,full} / N_{2,nightly} = (2n-1)^2$ such that the number of pixel additions is dominated by the nightly stacks, we can estimate the total number of pixel additions required by:

\begin{equation}
\begin{split}
N_{2,add} &\simeq N_{2,nightly}N_{pixels}N_{obs}n\\
& \simeq  \frac{4N_{1}N_{pixels}N_{obs}n}{(2n-1)^2}.
\end{split}
\end{equation}

Whereas for the 1-level grid, the number of pixel additions required is $N_{1,add} = N_{1}N_{pixels}N_{obs}n$. The resulting improvement in the number of required pixel additions is roughly

\begin{equation}
\frac{N_{1,add}}{N_{2,add}} \simeq \frac{(2n-1)^2}{4} \label{tree}
\end{equation}

So for a two-night arc, we estimate a factor of $\sim2.25$ improvement in the number of pixel additions, while for a three-night arc this increases to $\sim6.25$.

\begin{table*}[t]
\centering
\begin{tabular}{lcclc}
\multicolumn{5}{c}{\bf Table 1: Survey Parameters}\\
\hline
\hline
Survey \& Method &Baseline \& Tracking Error& Grid Limits & Spacing  & $N_{grid}$\\
\hline
Fraser et al. 2008 &$t_{b} \sim4$ hours&$1\arcsec.4 hr^{-1} < \dot{\theta} < 4\arcsec.1 hr^{-1}$ & $d\dot{\theta} = 0\arcsec.7hr^{-1} $ & 25\\
(searched by eye) &$\epsilon = 1\arcsec.6$, $F \simeq 0.57^{a}$ &$-10^\circ < \phi < +10^\circ$& $d\phi = 5^\circ$& \\
\hline
Fraser \& Kavelaars 2009 &$t_{b} \sim4$ hours &$0\arcsec.4 hr^{-1} < \dot{\theta} < 4\arcsec.5 hr^{-1}$ & $d\dot{\theta} = 0\arcsec.21hr^{-1} $& 95\\
(searched by eye) &$\epsilon =1\arcsec.25$, $F \simeq 0.61^{a}$& $-15^\circ < \phi < +15^\circ$& $d\phi = 7.5^\circ$&\\
\hline
Fuentes et al. 2009 &$t_{b} \sim8.5$ hours&$0\arcsec.7 hr^{-1} < \dot{\theta}_{\shortparallel} < 5\arcsec.1 hr^{-1}$ & $d\dot{\theta}_{\shortparallel} = 0\arcsec.1hr^{-1} $& 736\\
(automated detection)&$\epsilon =0\arcsec.6$, $F \simeq 0.76^{a}$&$-1\arcsec.4 hr^{-1} < \dot{\theta}_{\perp} < +1\arcsec.4 hr^{-1}$& $d\dot{\theta}_{\perp} = 0\arcsec.1hr^{-1} $&\\
&& $-15^\circ < \phi < +15^\circ$&&\\
\hline
 \end{tabular}
 \\
  {\footnotesize $^a$ : Maximum $S/N$ degradation factor due to tracking error, from Eqn. \ref{SNdeg}. $S/N' = F\times S/N_{0}$}
 \end{table*}

\begin{figure*}
\begin{centering}
\includegraphics[width=16cm]{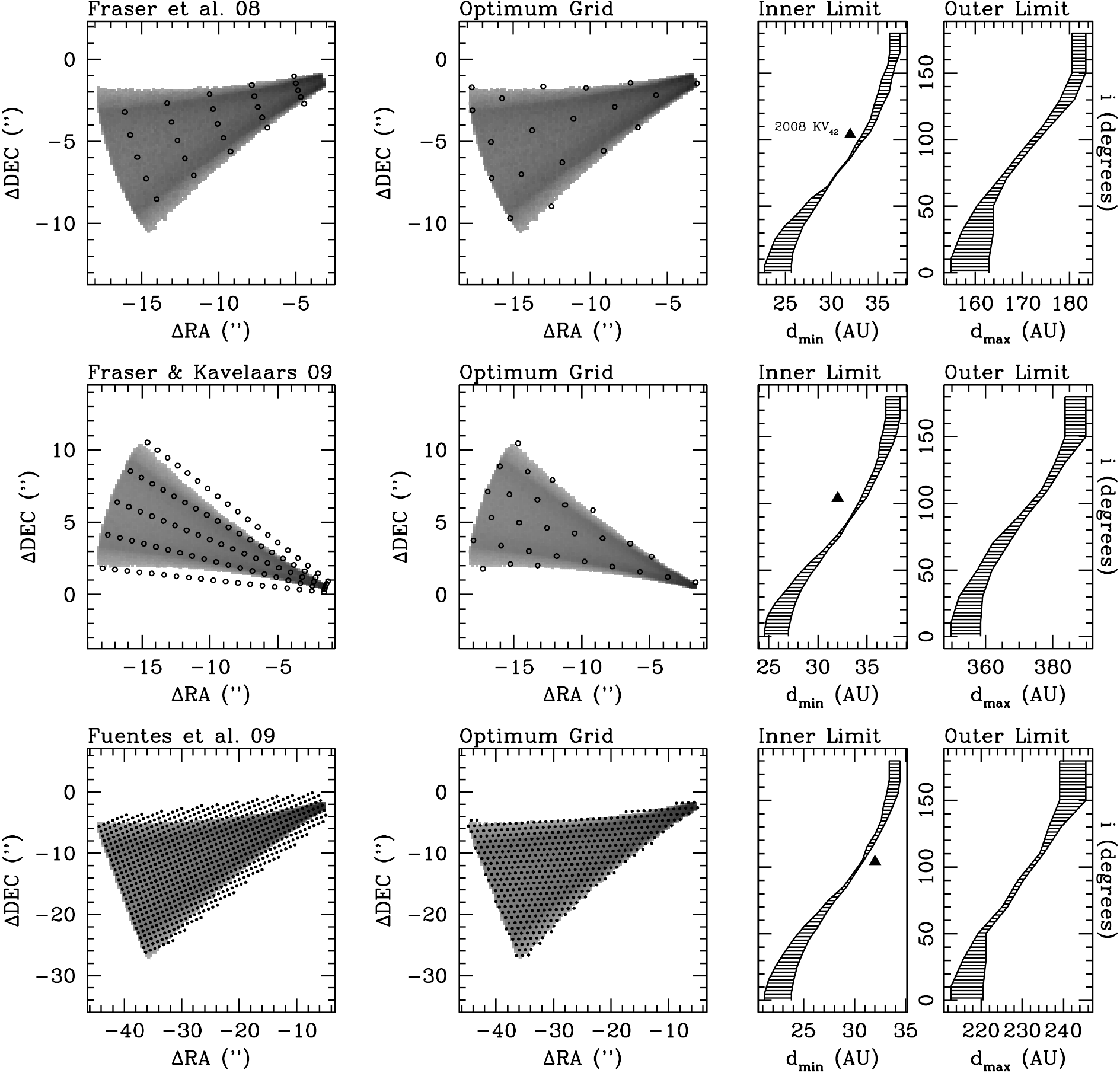}
\caption{Results from survey simulations. Top panels: Fraser et al. 2008. Middle panels: Fraser \& Kavelaars 2009. Bottom panels: Fuentes et al. 2009. Gray distributions represent density in ($d\alpha, d\delta$) of synthetic orbits generated as described in text. Left panels include overlay of each survey's search grid. Middle panels include overlay of our derived ``optimum'' search grid. Right panels show inner and outer limits in heliocentric distance $d$ vs. inclination $i$ derived for each survey. Cross-hatched region represents limit variation due to eccentricity. Black triangle represents the distance at discovery and inclination of 2008 KV$_{42}$.}
\label{All_surveys}
\end{centering}
\end{figure*}

\section{Comparison of Probabilistic Solution to Previous Survey Grids} \label{COMP}
In the following section, we will apply our probabilistic shift-vector generation method to previous surveys' observations, and compare the results to the methods used by the authors of the surveys. In order to determine the limits of parameter space searched by each survey, we generate our initial sample of synthetic orbits over very large ranges of heliocentric distance ($20-500$ AU), inclination ($0^\circ-180^\circ$), and eccentricity ($0-0.999$).  The sky motion of each synthetic orbit generated is then tested to ensure that it is within the ($\dot{\theta},\phi$) or ($\dot{\theta}_\shortparallel,\,\dot{\theta}_\perp$) ranges searched by the authors; as such, we only perform this characterization for surveys for which we can accurately reproduce the original search-grid from the literature. 

Three surveys are selected for comparison: Fraser et al. 2008, Fraser \& Kavelaars 2009, and Fuentes et al. 2009. These surveys are selected because they cover relatively large areas ($0.25-3$ square degrees) to relatively faint limits ($R\sim26-27$), and because they contain very complete information regarding their respective search grids and targeted orbital parameter space, which are outlined in Table 1. The comparisons between the original and optimized grids and the heliocentric distance range our analysis indicates each survey was sensitive to is listed in Table 2.

\begin{table*}
\centering
\begin{tabular}{lcccc}
\multicolumn{5}{c}{\bf Table 2: Survey Characterization}\\
\hline
\hline
Survey & $N_{opt}$& \% Improved & $d_{min}$ (AU)& $ d_{max}$ (AU)\\
\hline
Fraser et al. 2008 & 19 &24\%& $22-36$ & $164-184$\\
Fraser \& Kavelaars 2009 & 28 &  71\% & $24-37$ & $360-390$\\
Fuentes et al. 2009 & 494 & 33\% &$21-33$ & $220-245$\\
\hline
 \end{tabular}
 
 \end{table*}

\subsection{Fraser et al. 2008}

Fraser et al. 2008 searched $\sim3$ square degrees (taken over baselines ranging from 4 to 8 hours) for TNOs using fixed grids of rates and angles, searching the final stacks by eye. The images were acquired from several facilities, and detailed grid information is only supplied for the \textit{MEGAPrime} observations. These observations spanned $\sim4$ hours, and the search grid applied (described in Table 1) required 25 rates and angles be searched. The adopted search-grid spacing was fixed in $d\dot{\theta}$ and $d\phi$, resulting in a maximum tracking error as a function of $\dot{\theta}$ as described in Section \ref{AM}. We estimate that the resulting maximum tracking error was $\sim1\arcsec.6 \simeq 2.2\Gamma$, resulting in a maximum $S/N$ degradation factor of $F\simeq 0.57$. As the authors reported no sensitivity loss as a function of rate of motion, we adopt this as our uniform maximum tracking error. The authors state that their selection of rates and angles were designed to detect objects on circular orbits with heliocentric distances from $\sim25-100$ AU with inclinations as high as $70^\circ$.

Due to the short arc and large tracking error of this survey, our optimum grid (with 19 shift-vectors) is not radically more efficient ($\sim\!25\%$) than the original grid of 25 shift-vectors. The original and optimum grids, along with our initial sample, are illustrated in the left two top panels of of Figure \ref{All_surveys}.

Based on our simulation of this survey, we find that the minimum heliocentric distance the original grid is sensitive to is a strong function of inclination, with $d_{min} \simeq 22$ AU for $i = 0^\circ$ orbits, climbing to $d_{min} \simeq 30$ AU for $i = 70^\circ$.  This grid is in fact sensitive to inclinations as high as $180^\circ$ outside of 36 AU. The outer edge is similarly modified by inclination, but in all cases greater than the 100 AU goal: at the lowest, it is sensitive to objects at distances as high as 164 AU for $i \sim 0^\circ$, and at its highest it is sensitive to objects as distant as 184 AU for $i \sim 180^\circ$. These limits are illustrated in the top right panels of Figure \ref{All_surveys}.

\subsection{Fraser \& Kavelaars 2009}
\label{FraserSub}

Fraser \& Kavelaars 2009 searched $\sim0.25$ square degrees (taken over a $\sim4$ hour baseline) for TNOs with a grid of 95 rates and angles, aiming to be sensitive to objects on circular orbits with heliocentric distances from $\sim25-200$ AU. As in Fraser et al. 2008, the adopted search-grid spacing was fixed in $d\dot{\theta}$ and $d\phi$, and the final stacks were searched by eye. The authors state that they choose the ($d\dot{\theta},d\phi$) spacing to limit the maximum tracking error to $\sim2\Gamma$; we verify that at maximum $\dot{\theta}$ after 4 hours, the maximum tracking error of this search-grid is approximately $1.25 \arcsec \sim 1.8\Gamma$, resulting in a maximum $S/N$ degradation factor of $F\simeq 0.61$. As the authors reported no sensitivity loss as a function of rate of motion, we adopt this as our uniform maximum tracking error.

Based on our simulation of this survey, we find the original search-grid is over-ambitious as the high-$\phi$ regions of the grid are not populated by any real objects. No bound orbits observed on the ecliptic at opposition with heliocentric distance outside of $d > 28$ AU can have angles of motion as high as $15^\circ$ from the ecliptic (Eqn. \ref{maxangle}).  This, coupled with the oversampling at low-$\dot{\theta}$ relative to the maximum tracking error of $1.25 \arcsec$ resulted in excess shift-vectors compared to our optimum solution. The optimum grid requires 28 shift-vectors, which represents an improvement of over a factor of 3 compared the 95 shift-vectors searched by the authors. The original and optimum grids, along with our initial sample, are illustrated in the left two middle panels of of Figure \ref{All_surveys}.

Similar to Fraser et al. 2008, the minimum heliocentric distance the original grid is sensitive to is a function of inclination, with $d_{min} \simeq 24.5$ AU for $i = 0^\circ$ orbits, climbing to $d_{min} \simeq 31$ AU for $i = 70^\circ$.  This grid was also sensitive to inclinations as high as $180^\circ$ outside of 36.5 AU. The outer edge is significantly more distant than claimed: at the lowest, it is sensitive to objects at distances as high as 350 AU for $i \sim 0^\circ$, and at its highest it is sensitive to objects as distant as 385 AU for $i \sim 180^\circ$. These limits are illustrated in the middle right panels of Figure \ref{All_surveys}.

\subsection{Fuentes et al. 2009}
\label{FuentesSub}
Fuentes et al. 2009 searched $0.25$ square degrees (taken over a $\sim8$ hour baseline) for TNOs with a grid of 732 rates parallel and perpendicular to the ecliptic, aiming to be sensitive to objects with heliocentric distances from $\sim 20-200 AU$. The large number of resulting stacks necessitated the use of an automated detection pipeline. Their grid spacing was $ d\dot{\theta}_\shortparallel = d\dot{\theta}_\perp = 0\arcsec.1 \, hr^{-1}$ and motions were limited to $\pm15^\circ$ from the ecliptic. Over an $\sim8$ hour observational baseline, this grid spacing translates into a maximum tracking error of $\epsilon \simeq 0.6\arcsec \sim 0.8\Gamma$, resulting in a maximum $S/N$ degradation factor of $F\simeq 0.76$.

Like Fraser \& Kavelaars 2009, this survey also over-searches high-$\phi$ motions. Because of the small tracking error and non-optimum grid geometry, any small over-searched regions translate into a significant number of additional search vectors. The optimum grid requires 494 shift-vectors, which represents an improvement of $\sim\!33\%$ over the original search grid. The original and optimum grids, along with our initial sample, are illustrated in the left two middle panels of of Figure \ref{All_surveys}.

Similar to Fraser et al. 2008, the minimum heliocentric distance the original grid is sensitive to is a function of inclination, with $d_{min} \simeq 21.5$ AU for $i = 0^\circ$ orbits, climbing to $d_{min} \simeq 26.5$ AU for $i = 70^\circ$.  This grid was also sensitive to inclinations as high as $180^\circ$ outside of 33.5 AU. The outer edge is significantly more distant than claimed: at the lowest, it is sensitive to objects at distances as high as 220 AU for $i \sim 0^\circ$, and at its highest it is sensitive to objects as distant as 245 AU for $i \sim 180^\circ$. These limits are illustrated in the bottom right panels of Figure \ref{All_surveys}.

\section{Example application: LSST deep fields}
\label{LSST}



 \begin{figure*}
\begin{centering}
\includegraphics[width=8cm]{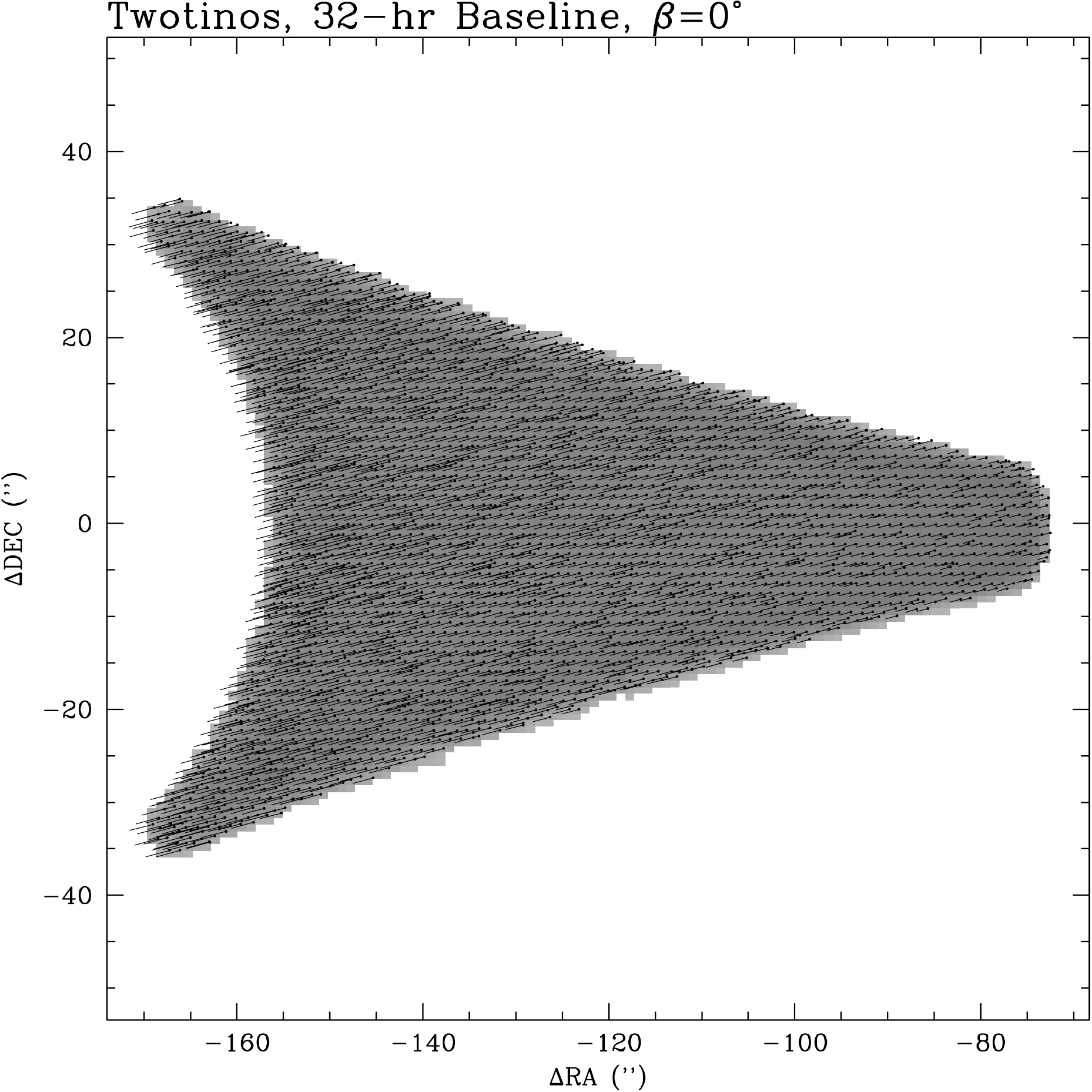}
\includegraphics[width=8cm]{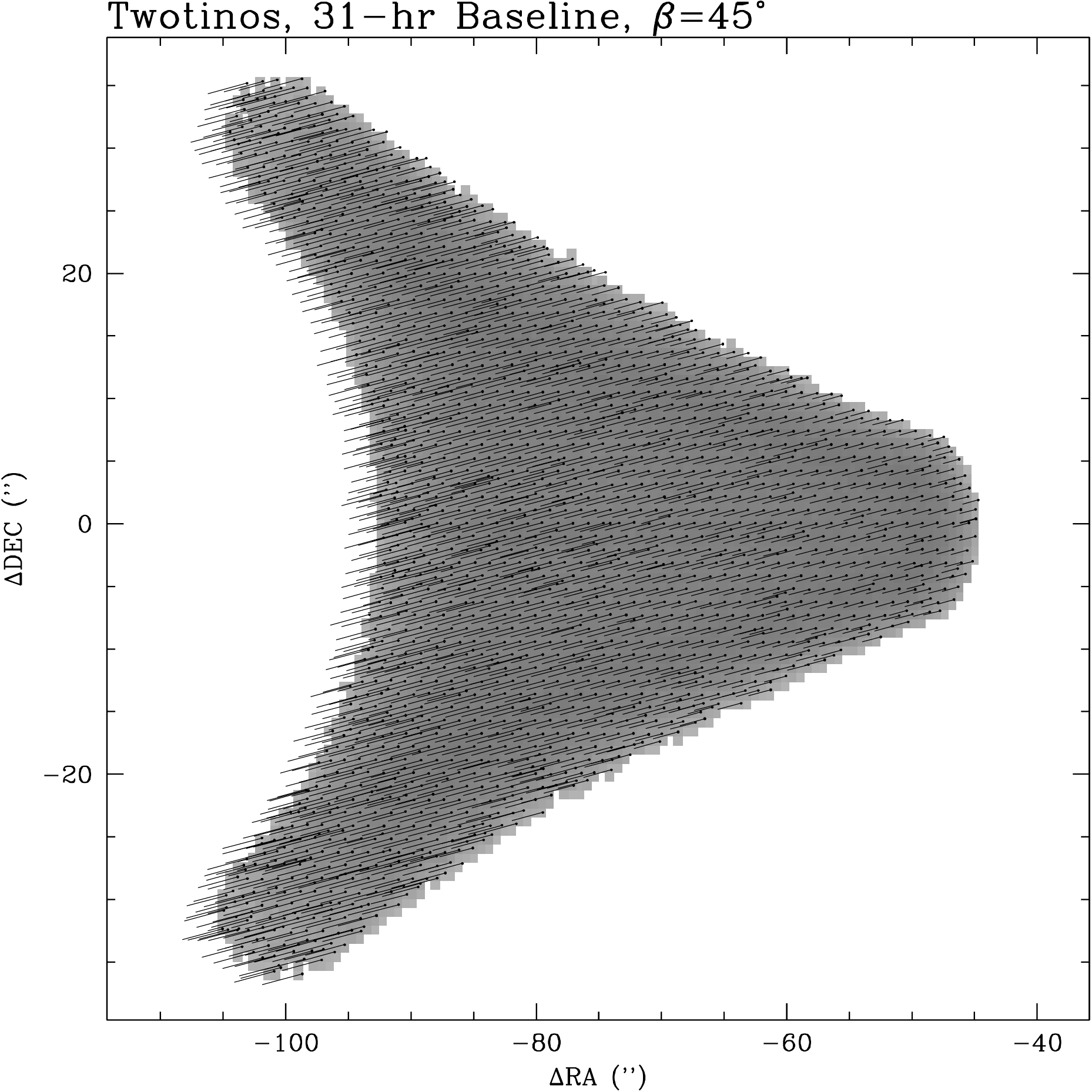}
\caption{Simulated motions of near-ecliptic Twotinos after two nights (gray distribution), optimum grid for LSST observations (points) and non-linear corrections to grid (lines; acceleration is nearly parallel to the celestial equator for the simulated fields, but a small angle has been added to separate the plotted lines from each other). Left panel: Opposition observations. Right panel: Observations at $45^\circ$ from opposition.}
\label{twotinos}
\end{centering}
\end{figure*}

A component of the LSST survey strategy will be to point to a single field and take hundreds of $\sim30$ second exposures, then later combine them in software to search for faint moving objects (Chesley et al. 2009). On a single winter night, a single 9.6 square degree field  at opposition can be observed for roughly 8 hours, resulting in 850 images with a combined depth of $r' \sim 28$. We will determine the feasibility of stacking multiple nights of data to search for specific TNO populations at even fainter magnitudes.

We have run simulations of similar LSST observations to demonstrate the utility of our method for generating shift-vectors. One of this method's chief advantages is the ability to strictly limit the orbital parameters of interest, being certain to sensitize the resulting stacks only to those orbits of interest without creating stacks at extraneous rates of motion. As such, the simulation described here is designed to detect a population that has some orbital parameters that are well-defined; namely, objects in the Neptune 2:1 resonance, or ``Twotinos.'' The orbital ranges used to simulate this population are listed in Table 3.

We compare observations made at opposition to those made at $45^\circ$ away from opposition, which have increasing contributions from non-linear components of motion. If an opposition field can be observed for 8 hours on a single night, fields at this elongation can be observed for $\sim \!7$ hours per night. We simulate observing both fields for one, two, and three nights, resulting in the total observational baselines listed in Table 4. The seeing is assumed to be the projected $75^{th}$-percentile in $r'$, roughly $0\arcsec.89$ (Tyson et al. 2009). We adopt $\epsilon_{max}=0.8\Gamma$, similar to what has been used in previous automated-detection-based pencil-beam surveys, resulting in $\epsilon_{max}\simeq0\arcsec.7$.

\begin{table}
\centering
\begin{tabular}{l|cc}
\multicolumn{3}{c}{}\\
\multicolumn{3}{c}{\bf Table 3: LSST Simulation Orbits}\\
\hline
\hline
\multicolumn{3}{l}{Twotinos}\\
\multicolumn{2}{c}{Parameter Range} & \multicolumn{1}{c}{Distribution} \\
\hline 
$a$ & 42.8 AU & Single-valued \\
$q$ & $25^{a}-42.8$ AU& Uniform \\
$e$ & $0-0.416$ & Uniform \\
$d$ & $25^{a}-60.6$ AU & $p(d) \propto d^{-2} $ \\
$i$ & $0^\circ - 45^\circ$ & Uniform \\
\hline
 \end{tabular}
 \end{table}

\begin{table*}
\centering
\begin{tabular}{lccccc}
\multicolumn{6}{c}{\bf Table 4: LSST Simulation Results}\\
\hline
\hline
\multicolumn{6}{l}{Twotinos}\\
$\beta$ & $t_{b} (hr) $ & Depth ($r'$ mag) & $N_{shift}$ & $N_{add}$ & $N_{search}$ \\
 \hline
$0^\circ$ & 8 & 28& 200 &$5.4\times10^{14}$ & $6.4\times10^{11}$\\
& 32 & 28.4& 3,359 & $1.8\times10^{16}$ & $1.1\times10^{13}$\\
 & 56 & 28.6& 7,743  & $6.4\times10^{16}$ & $2.5\times10^{13}$\\ 
 \hline
$45^\circ$ & 7 & 27.9& 113 &$2.7\times10^{14}$& $3.6\times10^{11}$\\
& 31 & 28.3& 1,965 &$9.4\times10^{15}$ & $6.3\times10^{12}$\\
 &  54 & 28.5 & 6,864 & $4.9\times10^{16}$& $2.2\times10^{13}$\\  
 \hline
 
 \end{tabular}
 \end{table*}

Table 4 contains the results of our simulations. We have estimated the required number of shift-vectors, pixel additions, and pixels to search (without the additional optimization from the tree of grids discussed in Section \ref{alias}). Figure \ref{twotinos} illustrates the optimum non-linear grids generated for both elongations. Also illustrated is the magnitude of the error between the final positions of real sources and the linear extrapolation of their position from their initial motion ($\dot{\theta}_0, \phi_0$). 

It is useful to compare the total required number of pixel additions ($N_{add}$) and pixels to search ($N_{search}$) to the most computationally-intensive digital-tracking survey to date. The HST survey performed by Bernstein et al. (2004) required $N_{add} \sim 10^{16}$ additions and $N_{search} \sim 7\times10^{13}$ pixels with 2004 computer technology. Because the LSST PSF is significantly larger than that of HST, fewer shift-vectors are required for the same observational baseline. In our LSST Twotino simulations, even the three-night opposition solution requires as many or fewer $N_{add}$ ($6.4\times10^{16}$ additions) and $N_{search}$ ($2.5\times10^{13}$ pixels). Including the potential improvement of a factor of $\sim6.25$ in the total number of pixel additions required with the addition of a second-level grid to the three-night arc (Eqn. \ref{tree}), it is clear that even with technology that will be over a decade out of date by the time LSST comes online, multi-night stacking of LSST data will be feasible for some TNO populations.  

Generalizing to the entire TNO population does not vastly increase the computational demand. We repeated the 8-hour opposition simulation allowing $a$ to vary uniformly over the same range as $d$ and the number of required shift-vectors increased by approximately 15\%. This increase is largely driven by the addition of low-eccentricity objects at low heliocentric distances.

\section{Conclusions}

We have shown that by generating search-grids for pencil-beam surveys in a probabilistic way and implementing an optimized grid geometry, significant reductions ($25-75\%$) can be made in the total number of operations required, and using a tree-of-grids structure can reduce the total pixel additions required even further. Besides improvements in computational requirements, this method also ensures uniform sensitivity to the entire targeted volume of orbital parameter space.

This method also provides a simple way to include non-linear components of motion necessary to track objects over long observational baselines. This combination of reduced computational requirements and simplicity of extending the observational arc makes the prospect of stacking multiple nights of data appear completely feasible. The additional depth and improved precision of measured orbital properties derived from multi-night stacks would be hugely beneficial if used in upcoming large-area surveys like LSST and Pan-STARRS. Simulations of the LSST deep observations show that, for certain TNO populations, it would require less computational effort to obtain the same depth as Bernstein et al. (2004) over 500 times the area by stacking multiple nights of LSST data with the methods described here. 

Additionally, this method allows for simple characterization of the orbital sensitivity of existing pencil-beam surveys in literature. From our characterization of the surveys by Fraser et al. (2008), Fuentes et al. (2009), and Fraser \& Kavelaars (2009), we have shown that the orbital sensitivity of these surveys varied from what was advertised. Their extra sensitivity to the motions of distant sources makes these surveys sensitive to populations like distant members of the Scattered Disk and Sedna-like objects, and their non-detection provide useful upper limits on these populations. We will explore these upper limits in future work. 

Since the discovery of extremely high-inclination, low-pericenter objects like 2008 KV$_{42}$ ($i \simeq 110^\circ$), it is important to understand these surveys' sensitivity to such populations. We note that Fuentes et al. (2009) is the only survey characterized here that was sensitive to objects with the same inclination and distance at discovery as 2008 KV$_{42}$ (see Figure \ref{All_surveys}, bottom-right panel), but we contend that it is unlikely that they would have recognized any detection as belonging to this retrograde population. Since follow-up has been performed only rarely for these deep surveys, orbital inclination and heliocentric distance usually remain degenerate, with the prograde solution lying at lower distances (though this degeneracy is rarely acknowledged). It is possible that detections labeled as high-inclination prograde objects may in fact be objects with retrograde orbits like 2008 KV$_{42}$ at greater distance. Only multi-night arcs or follow-up at later epochs can break this degeneracy and clearly identify objects belonging to rare TNO sub-classes.

\section{Acknowledgements}

Alex Parker is funded by the NSF-GRFP award DGE-0836694.

\end{document}